\documentclass{article}
\usepackage{emulateapj}
\usepackage{amsmath}
\usepackage{amssym}
\usepackage{psfig}
\usepackage{times}

\input{epsf}

\def\ms{\,{\rm ms}}

\def\t0{t_0}

\def\dd{{\rm d}}

\def\ms{\mu_s}

\newbox\grsign \setbox\grsign=\hbox{$>$} \newdimen\grdimen \grdimen=\ht\grsign
\newbox\simlessbox \newbox\simgreatbox \newbox\simpropbox
\setbox\simgreatbox=\hbox{\raise.5ex\hbox{$>$}\llap
     {\lower.5ex\hbox{$\sim$}}}\ht1=\grdimen\dp1=0pt
\setbox\simlessbox=\hbox{\raise.5ex\hbox{$<$}\llap
     {\lower.5ex\hbox{$\sim$}}}\ht2=\grdimen\dp2=0pt
\setbox\simpropbox=\hbox{\raise.5ex\hbox{$\propto$}\llap
     {\lower.5ex\hbox{$\sim$}}}\ht2=\grdimen\dp2=0pt
\def\simgt{\mathrel{\copy\simgreatbox}}
\def\simlt{\mathrel{\copy\simlessbox}}

\newcommand{\bez}{\begin{eqnarray*}}
\newcommand{\eez}{\end{eqnarray*}}
\newcommand{\be}{\begin{equation}}
\newcommand{\ee}{\end{equation}}
\newcommand{\beq}{\begin{eqnarray}}
\newcommand{\eeq}{\end{eqnarray}}
\newcommand{\bc}{\begin{center}}
\newcommand{\ec}{\end{center}}

\lefthead{BELOBORODOV} 
\righthead{MAGNETIC FLARES IN CYG X-1}
\begin{document}

%\title{Bulk motion in X-ray flares near accreting black holes}
%\title{Bulk motion in magnetic flares above black hole accretion disks}
%\title{Plasma ejection from magnetic flares above black hole accretion disks}
\title{Plasma ejection from magnetic flares and the X-ray spectrum of Cygnus 
X-1}
%\title{Plasma motion in magnetic flares and the X-ray spectrum of Cygnus X-1}
%\title{Plasma motion in magnetic flares above the accretion disk in Cyg X-1}
%\title{Plasma acceleration in magnetic flares and the X-ray spectrum of Cyg X-1}
%\title{Magnetic flares and the X-ray spectrum of Cyg X-1}
%\title{Plasma ejection from active regions of black hole accretion disks}

\author{Andrei M. Beloborodov\altaffilmark{1}} 
\affil{Stockholm Observatory, S-133 36, Saltsj{\"o}baden, Sweden;
andrei@astro.su.se} 
\altaffiltext{1}{Also at Astro-Space Center of Lebedev Physical Institute, 
Profsojuznaja 84/32, Moscow 117810, Russia}

\begin{abstract}

The hard X-rays in Cyg X-1 and similar black hole sources are possibly
produced in an active corona atop an accretion disk. We suggest that
the observed 
weakness of X-ray reflection from the disk is due to bulk motion of the 
emitting hot plasma away from the reflector.
A mildly relativistic motion causes aberration reducing X-ray emission towards 
the disk.  This in turn reduces the reprocessed radiation from the disk
and leads to a hard spectrum of the X-ray source.
The resulting spectral index is $\Gamma\approx 1.9/\sqrt{B}$ where 
$B=\gamma(1+\beta)$ is the aberration factor for a bulk velocity $\beta=v/c$.
%Both the spectral slope, 
The observed $\Gamma\approx 1.6$ and the amount of reflection, 
$R\approx 0.3$, in Cyg X-1 in the hard 
state can both be explained assuming a bulk velocity $\beta\sim 0.3$. 
We discuss one possible scenario: the compact magnetic flares are dominated by 
$e^\pm$ pairs which are ejected away from the reflector by the pressure of 
the reflected radiation. 
We also discuss physical constraints on the disk-corona model and argue that
the magnetic flares are related to magneto-rotational instabilities in the
accretion disk.

\end{abstract}


\bigskip

\section{Introduction}

%\medskip
 
Galactic black holes (GBHs) in their hard state and radio-quiet active 
galactic nuclei (AGNs) have similar X-ray spectra which are well explained 
by Comptonization of seed soft photons in a hot cloud with scattering 
optical depth $\tau_{\rm T}\sim 1$ and temperature $kT\sim 100$ keV (see
reviews by Zdziarski et al. 1997; Poutanen 1998). 
The Compton reflection feature in observed spectra and 
a fluorescent iron line indicate the presence of relatively cold gas 
in the vicinity of the X-ray source. The most likely reflector is an accretion 
disk, and the hard X-rays are possibly produced in a hot corona of the disk 
(e.g., Bisnovatyi-Kogan \& Blinnikov 1977; Liang 1979).
The dissipation of magnetic energy in the corona may feed the X-ray
luminosity (Galeev, Rosner, \& Vaiana 1979), 
while the underlying accretion disk reprocesses the incident X-rays
and supplies seed soft photons to the corona. 

A self-consistent
disk-corona model was developed, in which the coronal plasma is cooled by
its own radiation reprocessed in the disk (Haardt \& Maraschi 1993). 
The slope of the emerging X-ray spectrum is then determined by the fraction
of the X-ray luminosity which comes back to the source as a reprocessed
soft radiation. In a self-consistent model, this fraction equals $A^{-1}$
where $A$ is the Compton amplification factor of the hot corona. 
The resulting spectrum depends on the geometry of the corona. 
Ovservations favour a patchy corona consisting of separate active blobs
(Haardt, Maraschi, \&  Ghisellini 1994; Stern et al. 1995; Poutanen \&
Svensson 1996). 

The disk-corona model has recently been found to be
in conflict with observations of Cyg X-1 and similar black hole sources in 
the hard state (e.g., Gierli\'nski et al. 1997):
i) The observed spectrum is very hard which corresponds to a Compton 
amplification factor $A\simgt 10$ and implies soft photon starvation 
of the hot plasma (Zdziarski, Coppi, \& Lamb 1990). 
The model predicts $A\simlt 5$ unless the active blobs are elevated above 
the disk at heights larger than the blob size (Svensson 1996).
ii) The model with elevated blobs would yield a strong reflection component,
$R=\Omega/2\pi\approx 1$, where $\Omega$ is the solid 
angle covered by the cold matter as viewed from the X-ray source. 
The reported amount of reflection is small, $R\sim 0.3$. 

The photon starvation and weak reflection 
may be explained if the cold reflector is disrupted near the black hole. 
This would agree with the idea that accretion proceeds as a hot 
flow in the inner region (Shapiro, Lightman, \& Eardley 1976; 
Esin et al. 1998; Krolik 1998; Zdziarski 1998). 
One may fit the observed spectra with a toy model of inner hot cloud
upscattering 
%The inner hot flow may emit hard X-rays due to Comptonization of
soft photons supplied by the surrounding cold disk or by dense cloudlets 
inside the hot region 
%One may fit the observed spectra assuming such a geometry 
(e.g., Poutanen, Krolik, \& Ryde 1997; Zdziarski et al. 1998).

However, the weak reflection does not necessarily imply that the inner 
cold disk is disrupted. One suggested alternative is that the apparent
weakness of the reflection features may be due to a high ionization 
of the upper layers of the disk (Ross, Fabian, \& Young 1998). A detailed 
analysis remains to be done to check whether the highly ionized reflector
is consistent with observed spectra.

In this Letter, we suggest another alternative. We find that 
the disk-corona model may be reconciled with observations 
if the plasma in the active regions has a mildly relativistic 
bulk velocity directed away from the disk. 
We argue that the plasma heated in magnetic flares is likely to be 
ejected with such a velocity, especially if it is comprised of $e^\pm$ pairs.

\bigskip
 
\section{Magnetic flares in the corona} 

\medskip

The usually exploited model for the corona formation is that of 
Galeev et al. (1979, hereafter GRV). According to the model,
a seed magnetic field is exponentially amplified in the disk due to 
a combination of the differential Keplerian rotation and the turbulent 
convective motions. The amplification time-scale at a radius $r$ 
is given by $t_{\rm G}\sim r/3v$ where $v$ is the convective velocity. 
GRV showed that inside luminous disks the field is not able to dissipate 
at the rate of amplification, and buoyant magnetic loops must emerge 
above the disk surface where the magnetic field may annihilate quickly. 
The rate of magnetic energy production per unit area of the disk equals 
$F_B=2hw_B/t_{\rm G}$ where $w_B=B^2/8\pi$ is the magnetic energy density
and $h$ is the half-thickness of the disk.
Assuming that the magnetic stress $t_{r\varphi}=B_\varphi B_r/4\pi$ is 
responsible for the transfer of angular momentum in the disk, one can compare 
$F_B$ to the total surface dissipation rate, $F_+=3t_{r\varphi}c_s$, 
where $c_s$ is the sound speed in the disk (Shakura \& Sunyaev 1973). 
One then finds $F_B/F_+=h/r$ (taking into account that 
$2w_B/t_{r\varphi}=B_\varphi/B_r=c_s/v$ in the GRV model). 
Hence, the GRV mechanism is able to dissipate only a small fraction 
$\sim h/r\ll 1$ of the total energy. This is in conflict with the spectra of 
GBHs in the hard state, in which a large fraction of the energy is emitted in 
hard X-rays. 

Recent simulations of magnetic turbulence in accretion 
disks indicate that the magneto-rotational instability is likely 
to drive the turbulence (see Balbus \& Hawley 1998 for a review). This 
instability operates on a Keplerian time-scale which is typically $h/r$ times 
shorter than $t_{\rm G}$. It produces magnetic energy with rate $F_B\sim F_+$,
i.e., fast enough to explain the bulk of energy release as magnetic dissipation.
% of magnetic energy, as then $F_B\sim F_+$. 
Combined with the GRV argument for magnetic buoyancy, it follows that a large 
fraction of $F_+$ may dissipate in the corona.

The corona thus may be the place where magnetic stress driving the accretion is 
transported to and released. For a standard radiation pressure dominated disk,
$t_{r\varphi}\approx m_pc\Omega_{\rm K}/\sigma_{\rm T}$, where
$\Omega_{\rm K}$ is the Keplerian angular velocity. 
The magnetic field in the corona is likely to be strongly inhomogeneous and
it is plausible that the main dissipation occurs in localized blobs 
where the magnetic energy density, $w_B$, is much larger than $t_{r\varphi}$ 
given by the standard model.
The accumulated magnetic stress may suddenly be released on a 
time-scale $t_0\sim 10 r_b/c$ (the ``discharge'' time-scale, see Haardt 
et al. 1994), where $r_b$ is the blob size. 
This produces a flare of luminosity $L\sim r_b^3 w_B/t_0$, which 
may have a compactness parameter 
$l=L\sigma_{\rm T}/r_bm_ec^3\sim 10-10^3$.

The compact flares can get dominated by $e^\pm$ pairs created in $\gamma-\gamma$
reaction (e.g., Svensson 1986).
Pairs would keep an optical depth $\tau_{\rm T}\sim 1$ during the flare
and upscatter soft radiation entering the $e^\pm$ blob.
The blob temperature is typically in the range $50-200$ keV depending on $l$ 
and on whether there is a non-thermal $e^\pm$ tail. A strongly localized 
$e^\pm$ flare with $l\sim 10^2-10^3$ may have a low temperature, $kT<100$ keV, 
observed in some GBHs and AGNs. 
In the presence of a non-thermal $e^\pm$ tail, the flare may have even
$kT\simlt 50$ keV which is observed in the case of GX 339-4 
(Zdziarski et al. 1998). 

Consider now one possible mechanism of bulk acceleration which should
operate in an $e^\pm$ blob located above a reflector.
The blob luminosity is partly reflected, and hence the pair
plasma is immersed in an anisotropic radiation field. 
The net radiation flux, $F\sim L/r_b^2$, must efficiently accelerate 
the light pairs. The time-scale for acceleration to relativistic velocities is 
$t_a\sim m_ec/f\sim l^{-1}r_b/c$ where 
$f\sim F\sigma_{\rm T}/c$ is the accelerating radiative force. 
The shortness of $t_a\ll r_b/c$ implies that the bulk 
velocity saturates at some equilibrium value limited by the radiation drag
%which becomes important at relativistic velocities
(e.g., Gurevich \& Rumyantsev 1965; Sikora \& Wilson 1981).
We discuss the pair acceleration by reflected radiation in more detail in \S 4.

The bulk motion in the hot blob does not necessarily
mean that the blob itself moves. Instead, the dissipation region can be 
static. Then the escaping pairs are replaced by newly created $e^\pm$. 
The escaping pairs are immediately cooled down to the Compton temperature 
$kT_{\rm C}\sim 10$ keV.
% and form a 
%cold jet if they move along open magnetic field lines. If trapped by a closed 
%magnetic loop, the pairs accumulate and annihilate at the top of the loop. 
Note that in a bright $\gamma$-ray source, $\gamma-\gamma$ interactions would 
produce a lot of pairs above/between the active regions. 
An outflow of cold pairs covering the whole inner region of the disk may then
be created (Beloborodov 1998a). 

\bigskip

\section{Bulk motion and reflection}

\medskip

It was noted previously that bulk motion of the flaring plasma may reduce 
the reflected radiation from an accretion disk (Wo\'zniak et al. 1998). 
Reynolds \& Fabian (1997) calculated the impact of motion on the Fe K$\alpha$ 
line. Here, we evaluate the amount of reflection and study how  
fast motion affects the coupling between the X-ray source and the reflector. 
We assume that the active blob luminosity is due to Compton amplification of 
the reflected (reprocessed) radiation, and the multiply upscattered photons get
isotropized in the plasma comoving frame. The angular distribution of the blob 
luminosity in the lab frame is then given by
\begin{equation}
   L(\mu)=\frac{L}{2\gamma^4(1-\beta\mu)^3},
\end{equation}
where $\mu=\cos\theta$, $\theta$ is the angle between the ray and 
the plasma velocity, $\gamma=(1-\beta^2)^{-1/2}$, and  
$L=\int_{-1}^1 L(\mu)\dd\mu$ is the total luminosity of the blob.
For a typical spectral index, $\Gamma\sim 2$, the relativistic 
transformation of the specific luminosity, $L_\nu(\mu)$, is the same as
that of the bolometric luminosity (Rybicki \& Lightman 1979).
In equation (1), we assume that the dissipation {\it region} is static
and there is no 
retardation effect (cf. Rybicki \& Lightman 1979). For a static blob with a 
life-time $t_0\gg r_b/c$ one may consider a 
time-independent picture in a fixed geometry like the usual disk-corona model. 
The only difference is that now the hot plasma in the blob has a bulk velocity.

We will assume that the velocity is perpendicular to the disk.
The X-ray luminosity striking the disk is then given by 
\begin{equation}
 L_-=\int_{-1}^0 L(\mu)\dd\mu=\frac{L(1+\beta/2)}{2\gamma^4(1+\beta)^2}.
\end{equation}
For a system inclination $\mu=\cos\theta$, the apparent strength of reflection, 
$R=\Omega/2\pi$, is
\begin{equation}
  R=\frac{L_-}{L(\mu)}=\frac{(1+\beta/2)(1-\beta\mu)^3}
                     {(1+\beta)^2}.
\end{equation}

The reprocessed radiation from the disk is partly intercepted by the active 
blob. The intercepted soft luminosity may be expressed in terms of
an effective $\mu_s$,   
\begin{equation}
  L_s=\chi\int_{-1}^{-\mu_s} L(\mu)\dd\mu.
\end{equation} 
The case $\mu_s=0$ describes a slab geometry of the active region, 
while $\mu_s\sim 1/2$ roughly corresponds to a blob with radius of order 
its height. The factor $\chi=1-a$ represents the efficiency of reprocessing,
%reflected radiation which is reprocessed into soft photons, 
$a\approx 0.1-0.2$ is the disk albedo (e.g., Magdziarz \& Zdziarski 1995).

The fraction of $L$ which returns as soft radiation
is a function of $\beta$ and $\mu_s$, 
\begin{equation}
  \frac{L_s}{L}=\frac{L_{s0}}
                      {L}\frac{1+(1+\mu_s)\beta/2}
                      {\gamma^4(1+\beta)^2(1+\mu_s\beta)^2},
\end{equation} 
where $L_{s0}=\chi(1-\mu_s)L/2$ corresponds to the usual 
case of a static corona ($\beta=0$). 
With increasing $\beta$, the intercepted soft luminosity is suppressed.

The spectral index of the Comptonized radiation emerging from the blob 
is determined by the Compton amplification factor taken in the plasma comoving 
frame. 
A luminosity ${\rm d}L=L(\mu){\rm d}\mu$ transforms into the comoving frame as 
${\rm d}L^c=\gamma^2(1-\beta\mu){\rm d}L
=(1+\beta\mu^c)^{-1}{\rm d}L$, 
where index $c$ stands for the comoving frame (see Rybicki \& Lightman 1979). 
This gives $L^c=L$ and $L_s^c=\gamma^2(L_s-\beta\Phi_s)$, where 
$\Phi_s=\int L_s(\mu)\mu {\rm d}\mu$ is the 
total flux of the soft radiation through the blob.
We approximate $L_s(\mu)=L_s/(1-\mu_s)$
at $\mu>\mu_s$ and $L_s(\mu)=0$ at $\mu<\mu_s$. Then
\begin{equation}
  \Phi_s=\frac{L_s(1+\mu_s)}{2},
\end{equation}
which yields $L_s^c/L_s=\gamma^2[1-\beta(1+\mu_s)/2]$.
We then get the amplification factor
\begin{equation}
 A\equiv\frac{L}{L_s^c}=
  \frac{2\gamma^2(1+\beta)^2(1+\mu_s\beta)^2}
       {\chi(1-\mu_s)[1-(1+\mu_s)^2\beta^2/4]}.
\end{equation}
For $\ms\simgt 0.3$, this expression can be approximated as 
$A\approx A_0B^\psi$ within 1\% accuracy in the range $-0.1<\beta<0.7$.
Here, $A_0=2/\chi(1-\ms)$, $B\equiv\gamma(1+\beta)$, and $\psi=2.7$, $2.9$, and 
$3.1$ for $\ms=0.3$, $0.4$, and $0.5$, respectively.

\bigskip

\section{Bulk velocity in a pair-dominated flare}

\medskip

We now estimate the expected bulk velocity for a pair-dominated flare  
using a simple toy model.
The power consumed by the $e^\pm$ blob is the sum of two parts: the primary 
injected power due to dissipation of magnetic energy, 
$L_{\rm diss}\approx L-L_s$,
and the reflected radiation intercepted by the blob, $\approx L_s$.
$L_{\rm diss}$ is probably injected in the 
form of relativistic particles which are supposed to share their energy 
and momentum with the thermal $e^\pm$ plasma, most likely due to collective 
effects or due to synchrotron self-absorption (Ghisellini, Guilbert, \& 
Svensson 1988). Let $\Phi_{\rm diss}$ be the total vertical flux of the 
injected energy. The injected particles may accelerate or decelerate the 
thermal plasma depending on the angular distribution of injection.
The reflected radiation always tends to accelerate the coronal plasma away from
the disk. 

The energy streaming into the blob per unit time equals 
$L_{\rm diss}+L_s=L$, and the corresponding net energy flux equals 
$\Phi_{\rm diss}+\Phi_s=\Phi$. The energy consumed during time $\dd t$
increases the blob momentum by $\dd p=(\Phi/c){\rm d}t$ and the inertial mass
by $\dd m=(L/c^2)\dd t$. The net acceleration is 
$\dd\beta=\dd(p/mc)=(\Phi-\beta L)\dd t/mc^2$.
The $e^\pm$ plasma  
accelerates on a short time-scale (see \S 2), 
and the bulk velocity should relax to an equilibrium value for which the net
acceleration vanishes. The equilibrium velocity is determined by the equation 
\begin{equation}
  \Phi-\beta L=0.
\end{equation}
This equation also expresses the condition that the total energy flux 
vanishes in the comoving frame, $\Phi^c=\gamma^2(\Phi-\beta L)=0$. 
%The problem is 
%formally equivalent to the problem of equilibrium motion of a blob absorbing 
%radiation $L$ with a total flux $\Phi$. This equivalence is due to the 
%assumption that the injected particles are relativistic and their flux 
%transforms into the comoving frame like a radiation flux. 

We do not possess a detailed model of injection in the blob and 
do not know the feedback of the plasma bulk velocity 
on the angular distribution of the injection. Therefore, we
consider two particular cases: 

\medskip

\noindent
{\it (a) $L_{\rm diss}$ is isotropic in the lab frame}

Then $\Phi_{\rm diss}=0$, and the equilibrium condition (8) combined
with (6) yields the relation
\begin{equation}
   \frac{2\beta}{1+\mu_s}=\frac{L_s}{L}.
\end{equation}
On the other hand, $L_s/L$ is determined by equation (5). Equating 
(5) and (9), we get an equation for the self-consistent bulk velocity
and find $\beta\sim 0.1$ for $0\leq\ms\simlt 0.5$. 

\medskip

\noindent
{\it (b) $L_{\rm diss}$ is isotropic in the comoving frame}

In this case, the energy flux associated with $L_{\rm diss}$ vanishes in the 
comoving frame and the equilibrium velocity is determined by the soft
radiation only, $\beta=\Phi_s/L_s$. This yields $\beta=1/2$ for 
$\mu_s=0$, and $\beta=3/4$ for $\mu_s=1/2$.

\medskip

A proton fraction exceeding $m_e/m_p$ would increase the plasma 
inertia, and then the bulk velocity may fall below the equilibrium value. 
Pairs will tend to stream through the heavy proton component
and a plasma instability may be initiated.
In the stationary case, the stream velocity establishes itself just at the 
instability threshold which may be expected to be
comparable with the thermal electron velocity, $\sim c/2$.

\medskip

\section{Conclusions and Discussion}

\medskip

The hard state of accreting black holes may be explained as a state in which
a large fraction of the luminosity is released in a magnetic corona atop a 
cold accretion disk.
The magnetic energy is probably generated by the magneto-rotational instability
in the disk and then transported to the corona due to buoyancy.
%This implies that the disk magnetic field is amplified on
%a Keplerian time-scale and transported to the corona due to buoyancy.
%The mechanism which can account for such a fast generation of magnetic energy
%is the magneto-rotational instability. 
The soft state may be observed when the coronal activity is suppressed and 
most of the energy is dissipated inside the disk.

The energy release in the corona is likely to proceed in compact bright flares, 
where the local radiation flux strongly exceeds the average surface flux from 
the disk. If the flaring plasma is comprised of $e^\pm$ pairs, it should 
be accelerated away from the disk by the pressure of the reflected radiation.
The bulk velocity is then expected to be in the range 
$0.1\simlt\beta\simlt 0.7$. 

The radiative acceleration of a pair plasma is one possible reason for
bulk motion in the flares. Even if the flare is dominated by a normal 
proton plasma, the heating may be accompanied by pumping a net momentum into 
the hot plasma at a rate $\sim L/c$. The transferred momentum per particle 
per light-crossing time, $r_b/c$, is $\sim l m_ec$. 
As the flare duration $t_0\gg r_b/c$, the protons may acquire a momentum 
comparable to $m_pc$. Note that magnetic flares where hot plasma is 
ejected towards the disk are possible. This case is formally described by 
equations (1-7) with $\beta<0$.

The impact of the source velocity on the observed reflection, $R$ (eq. [3]), 
and the Compton amplification factor, $A$ (eq. [7]), is summarized in Figure 1.
Comptonization in the source produces a power-law X-ray spectrum.
The photon spectral index, $\Gamma$, is related to $A$ by 
\begin{equation}
   \Gamma\approx 2.33 (A-1)^{-\delta},
\end{equation}
where $\delta\approx 1/6$ for GBHs and $\delta\approx 1/10$ for AGNs.
This simple formula approximates within a few percent the results of 
calculations we performed by using the code of 
Coppi (1992) (see Beloborodov 1999 for details). Combined with equation (7),
this yields the dependence $\Gamma(\beta)$. 
Assuming a typical $\ms\sim 0.5$, one can approximate $\Gamma(\beta)$ as
$\Gamma\approx 1.9B^{-0.5}$ for GBHs and $\Gamma\approx 2B^{-0.3}$ for 
AGNs, where $B=\gamma(1+\beta)$.

From the amount of reflection in Cyg X-1, $R\sim 0.3$, we infer a bulk velocity
of the emitting plasma $\beta\approx 0.3$ for a system
inclination $\sim 30^{\rm o}$ (see the bottom panel in Figure 1). 
On the other hand, from the reported intrinsic slope $\Gamma\sim 1.6$ and
the corresponding $A\sim 10$, we again infer the same $\beta\approx 0.3$.
Thus, $\beta\sim 0.3$ resolves the two problems of the disk-corona model 
mentioned in the Introduction. 
The beaming of the corona luminosity, $L_X$, also leads to the weakness
of observed reprocessed blackbody component in the spectrum, 
$L_{\rm bb}\sim \chi R L_X\sim L_X/4$.
Note that the inferred bulk velocity is comparable with the thermal electron
velocity, $\sim c/2$ for $kT\sim 100$ keV. 

The ejection model may apply to both GBHs and AGNs. 
The velocity of ejected hot plasma, $\beta$, may vary (and in some cases it 
might be $\beta<0$). An increase in $\beta$ leads to decreasing $R$ and 
$\Gamma$. A correlation between $R$ and $\Gamma$ is observed in the spectra
of GBHs and AGNs (Zdziarski, Lubi\'nski, \& Smith 1999) and it is well  
reproduced by our model (Beloborodov 1999).
Note also that fast outflows may manifest themselves in optical polarimetric
observations of AGNs (Beloborodov 1998b).

\bigskip

I am grateful to J. Poutanen and R. Svensson for many helpful discussions
and to the referee, A.~A. Zdziarski, for valuable remarks. I also thank
A.~C. Fabian and J.~H. Krolik for comments.
This work was supported by the Swedish Natural Science Research Council and 
RFFI grant 97-02-16975. 

\smallskip
\centerline{
\epsfxsize=10cm {\epsfbox{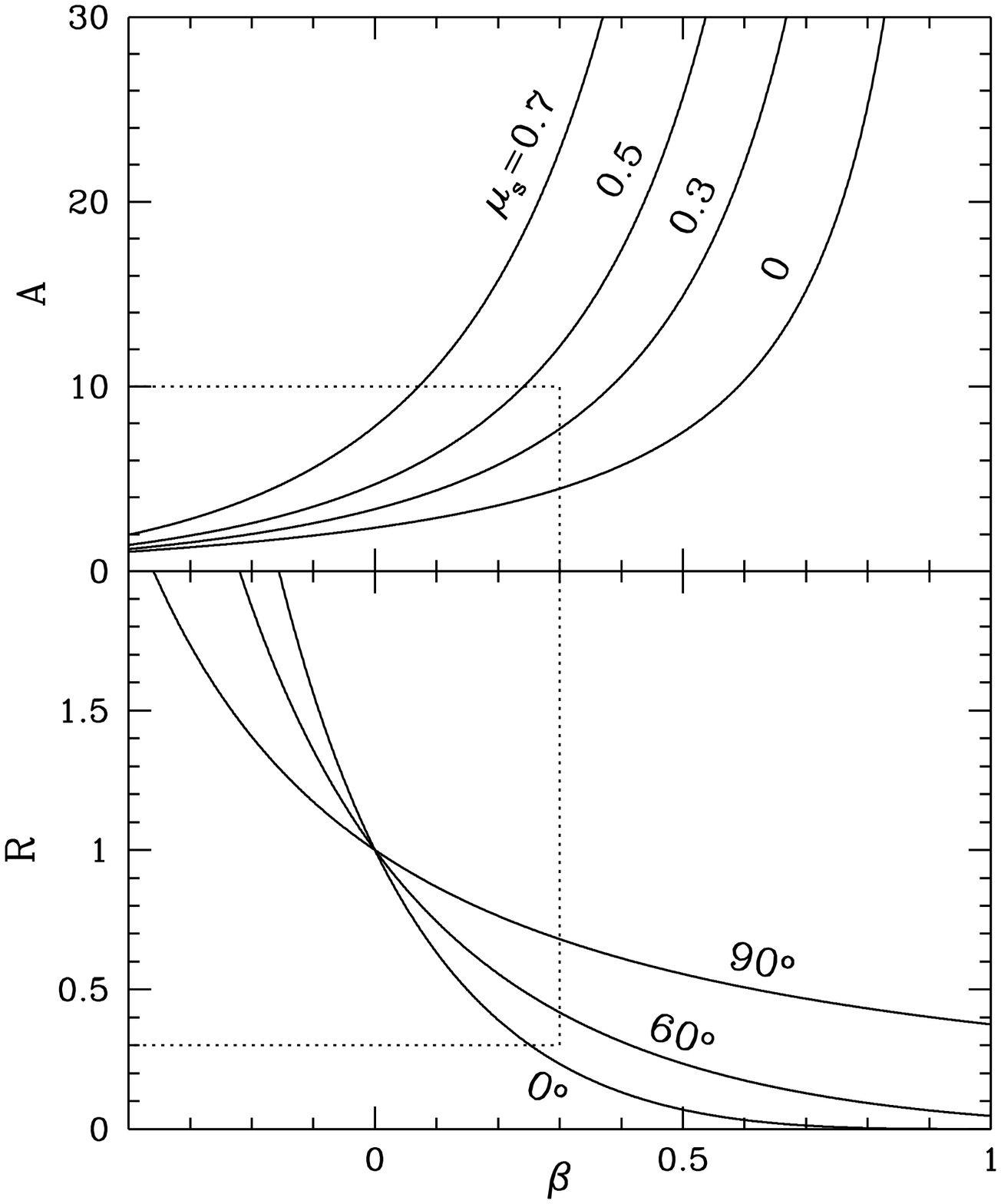}}
}
%\bigskip 
\figcaption {
{\it Top panel:} The Compton amplification, $A$, of the reprocessed radiation 
as a function of the plasma bulk velocity, $\beta$, (see eq. [7]). 
A reprocessed fraction $\chi=0.85$ is assumed. The 
parameter $\ms$ describes the flare geometry (see eq. [4]).
{\it Bottom panel:} The apparent amount of reflection, $R$, for the disk 
inclinations $\theta=0$, $60^{\rm o}$, and $90^{\rm o}$ (see eq. [3]). 
The observed $R$ may be further reduced as a result of the scattering
of the reflected radiation in the blob.
The dotted line shows the expected A and R for $\beta=0.3$. 
\label{fig1}}

\end{document}